\documentclass{emulateapj}

\newcommand{\um}{$\mu$m~}
\newcommand{\ums}{$\mu$m}

\shorttitle{High Redshift Obscured Sources}
\shortauthors{Houck et al.}

\begin{document}

\title{Spectroscopic Redshifts to z $>$ 2 for Optically Obscured Sources
  Discovered with the Spitzer Space Telescope }

\author{J. R. Houck\altaffilmark{1}, B. T. Soifer\altaffilmark{2}, D.
  Weedman\altaffilmark{1}, S. J. U. Higdon\altaffilmark{1}, J. L.
  Higdon\altaffilmark{1}, T. Herter\altaffilmark{1}, M. J. I. Brown\altaffilmark{3,4}, A.
  Dey\altaffilmark{4}, B. T. Jannuzi\altaffilmark{4}, E. Le
  Floc'h\altaffilmark{5}, M. Rieke\altaffilmark{5}, L.
  Armus\altaffilmark{2}, V. Charmandaris\altaffilmark{1} B. R.
  Brandl\altaffilmark{7}, \& H. I. Teplitz\altaffilmark{2}}

\altaffiltext{1}{Astronomy Department, Cornell University, Ithaca, NY 14853; jrh13@astro.cornell.edu}
\altaffiltext{2}{Spitzer Science Center, California Institute of Technology, 220-6, Pasadena, CA 91125}
\altaffiltext{3}{Department of Astrophysical Sciences, Princeton University, Peyton Hall, Princeton, NJ 08544-1001}
\altaffiltext{4}{National Optical Astronomy Observatory, Tucson, AZ 85726}
\altaffiltext{5}{Steward Observatory, University of Arizona, Tucson, AZ 85721}
\altaffiltext{6}{Leiden Observatory, 2300 RA Leiden, The Netherlands}

\begin{abstract}
  
  We have surveyed a field covering 9.0 degrees$^{2}$ within the NOAO
  Deep Wide-Field Survey region in Bo\"{o}tes with the Multiband
  Imaging Photometer on the Spitzer Space Telescope (SST) to a
  limiting 24 $\mu$m flux density of 0.3 mJy.  Thirty one sources from
  this survey with F$_{\rm 24\,\mu m}$ $>$ 0.75\,mJy which are
  optically very faint ($R$ $\ga$ 24.5\,mag) have been observed with
  the low-resolution modules of the Infrared Spectrograph on
  SST. Redshifts derived primarily from strong silicate absorption
  features are reported here for 17 of these sources; 10 of these are
  optically invisible ($R$ $\ga$ 26\,mag), with no counterpart in
  $B_W$, $R$, or $I$.  The observed redshifts for 16 sources are 1.7
  $<$ z $<$ 2.8.  These represent a newly discovered population of
  highly obscured sources at high redshift with extreme infrared to
  optical ratios. Using IRS spectra of local galaxies as templates, we
  find that a majority of the sources have mid-infrared spectral
  shapes most similar to ultraluminous infrared galaxies powered
  primarily by AGN. Assuming the same templates also apply at longer
  wavelengths, bolometric luminosities exceed $10^{13}$L$_{\odot}$.

\end{abstract}



\section{Introduction}

Opening the infrared wavelength regime for discovery is a primary
objective of the final Great Observatory, the Spitzer Space Telescope
\citep{wer04}. It was anticipated that new categories of sources would
be revealed in various surveys.  It was also anticipated that some of
these sources would be very faint optically, as dust obscuration is
probable in infrared-luminous sources, so the Infrared Spectrograph on
Spitzer (IRS\footnote{The IRS was a collaborative venture between
Cornell University and Ball Aerospace Corporation funded by NASA
through the Jet Propulsion Laboratory and the Ames Research Center} --
Houck, et al. 2004) was designed to include low resolution modules
which could determine redshifts for dusty sources. High-redshift,
optically faint but infrared-loud galaxies were first hinted at by the
deepest IRAS surveys \citep{hou84}, and partially uncovered in deep
surveys with ISO (Dole et al. 2001, Elbaz et al. 2002). The
sensitivity and wavelength coverage of the IRS enables us to measure
the redshifts and power sources within such galaxies even if they are
too faint for optical spectroscopy.

In order to search for such optically-faint infrared sources, we have
surveyed the Bo\"{o}tes field of the NOAO Deep Wide-Field Survey
(NDWFS, \citet{jan99}) with the Multiband Imaging Photometer for
Spitzer (MIPS; \citet{rie04}) to produce a catalog of mid-infrared
sources in 9.0\,deg$^{2}$ to flux density limits at 24\,\um of
0.3\,mJy.  This field was chosen because the deep and well calibrated
optical imagery in $B_W$, $R$, and $I$ bands makes possible the
identification of infrared sources with very faint optical
counterparts and confident selection of infrared sources lacking
optical counterparts to very deep limits.  Our overall objective for
the current study was to select the infrared sources which are
faintest optically but bright enough for feasible redshift
determination with the IRS, taking a limit of 0.75\,mJy for these
initial observations.

Details of the survey will be reported elsewhere.  We used the MIPS
and optical surveys to select and inspect all sources with 24\,$\mu$m
flux densities above 0.75\,mJy and with $I$ $\ge$ 24\,mag. Of the 4273
MIPS 24\,\um sources brighter than 0.75\,mJy, 114 met this optical
magnitude criterion.  Of these, 27 sources have no optical
identification in any band, meaning they are fainter than about
26.5\,mag in $B_W$, 26\,mag in $R$, and 25.5\,mag in $I$ (all Vega
magnitudes).  Based on comparison of positions between the many
identified optical counterparts and infrared sources, the formal
5\,$\sigma$ positional criterion for non-identification of 24\,\um
sources with flux densities above 0.75\,mJy is a distance $>2\arcsec$
from the nearest optical source.  We also rejected infrared sources
close to a bright galaxy or within clusters of faint galaxies with the
presumption that redshifts of the infrared sources can be determined
by association with the optical sources.  As a result of these
additional criteria, our initial target list contained 17 sources with
no optical identifications, and all of these have been observed with
the IRS. We also observed an additional 13 sources having an optical
counterpart but fainter than 24.5\,mag in $R$. Finally, one source was
included (\#13 in Table 1) because it is relatively bright at 24\,\um
but has an extreme IR/optical ratio (2.3\,mJy at 24\,\um and
$R$=23.8\,mag).  In this paper, we report redshifts, physical
properties and luminosities for 17 of the 31 sources which show
sufficiently strong spectral features that redshift measurements can
be made with the IRS.

\section{Observations and data reduction}

Observations and results for the sources discussed in this paper are
summarized in Table 1, giving the name (with full coordinates), the
24\,\um flux density, optical magnitudes of the MIPS 24\,\um source if
detected, integration times for the IRS observations, redshift
determinations, and other source characteristics. The MIPS data were
obtained with an effective integration time of $\sim$90\,sec per sky
pixel and reduced with the MIPS Data Analysis Tool
\citep{gor04}. Point source extraction was performed using an
empirical point spread function (PSF) constructed from the brightest
objects found in the 24\,$\mu$m image and subsequently fitted to all
the sources detected in the data. We allowed for a multiple match
fitting to deal with blended cases. We finally derived the flux of
each object using the scaled fitted PSF after applying a slight
correction to account for the finite size of the modeled point spread
function.

Spectroscopic observations were made with the IRS Short Low module in
order 1 only (SL1) and with the Long Low module in orders 1 and 2 (LL1 and
LL2), described in \citet{hou04}.  These give low resolution spectral
coverage from $\sim$8\,\um to $\sim$35\,\ums.  Sources were placed on
the slits by offsetting from nearby 2MASS stars. 
Observed images were processed with version 10.5 of the SSC pipeline.
Background subtraction and extraction of
source spectra along the slit was done with the SMART analysis package
\citep{hig04b}.  All spectra discussed in this paper are shown in
Figures~1 and 2.  

\section{Determining redshifts of sources}

We determine redshifts by using bright, local ``template'' sources for
which we have full spectral coverage with the IRS.  In almost all
cases, a redshift fit depends on the presence of a strong decrease in
the continuous spectrum beginning around $9\,\mu$m in the source rest
frame (for example, see \citet{hig04a}).  This drop is identified with
the strong silicate absorption feature centered at 9.7 \um; this
feature is common in ultraluminous infrared galaxies (ULIRGs) at low
redshift (e.g. Armus et al. 2004).  An extreme case of this type of
absorption is seen in the IRS spectrum of the ULIRG IRAS F00183-7111
\citep{spo04a}.  Strong spectral emission features, such as the
7.7\,\um and 8.6\,\um PAH features, can be present without the
presence of deep silicate absorption -- e.g., in the unobscured
starburst nucleus of NGC\,7714 \citep{bra04}.  The well known
low-redshift ULIRGs Arp\,220\,and Mrk\,231 are intermediate cases.
Arp\,220 has deep silicate absorption and strong PAH emission
\citep{spo04b}, while Mrk\,231 has weak silicate absorption, but
little or no PAH emission.  To cover the range from pure absorption to
pure emission, we have fit all spectra with the IRS templates of
NGC\,7714, F00183-7111, Arp\,220, and Mrk\,231. (The IRS spectra of
Arp 220 and Mrk 231 will be discussed in more detail in future
papers.)  A formal $\chi ^2$ routine gives the optimal redshift fit
for each template.  Once such fits were formally determined, the four
were compared to estimate which one gave overall the best fit to the
observed spectrum.  That is the result listed in Table~1 and shown in
Figure~1.

Figure~2 illustrates the effects of alternative templates for our
brightest 24\,\um source, \#9 in Table 1, showing that the redshift
uncertainty for a given source primarily depends on source S/N and
strength of the features rather than choice of template.  For the
sources in Table 1, this uncertainty typically is $\pm$ 0.2 in z; the
formal uncertainty derived from the $\chi ^2$ fit for the template
adopted is given for each source in Table 1.

The 17 sources out of our 31 observed spectra with measured redshifts
derived from template fits are all shown in Figure~1.  The 14
remaining sources do not have sufficiently strong features to list a
spectroscopic redshift.  An example of one of these sources is also
shown in Figure~1 to illustrate the obvious difference between spectra
with and without strong mid-infrared features.  Of the 17 sources, 15
have unambiguous redshifts based on the $\chi ^2$ fits shown, and all
of these have $1.7 < z < 2.8$.  The two sources for which we have less
confidence in the redshift include the only source, \#11, with
assigned z $<$ 1. The redshift of 0.70 arises because all three of the
templates containing absorption give a better fit at this redshift
than at the higher redshifts characterizing all of the other sources;
however, the spectrum beyond 30\,\um for this source does not have
sufficient S/N to show the drop in flux density which would be present
for a strong absorption feature at high redshift.  Source \#2 is the
only source for which the template fits produce an ambiguous redshift
based on $\chi ^2$.  Comparably good fits arise from Arp\,220 at z =
0.57 or NGC\,7714 at z = 1.86.  The NGC\,7714 fit is adopted because
of our presumption that a dusty source is more likely to be optically
faint at high redshift, but if this fit is correct, this is the only
source with a pure PAH emission spectrum.
 
\section{Discussion}

These results clearly demonstrate the presence of a population of
optically-obscured infrared sources at redshifts 1.7 $<$ z $<$
2.8. There are significant selection effects in our result that
constrain the redshift range in which we can identify this obscured
population.  We have deliberately selected against optically visible
sources, which generally forces sources to significant redshift,
probably z $\ga$ 1, because the rest-frame UV is most affected by
internal obscuration.  For 1.0 $\la$ z $\la $1.6, any deep 9.7 \um
silicate absorption feature in the spectrum would significantly affect
survey detections at 24\,$\mu$m, so absorbed objects would selectively
fall out of the 24\,\um sample.  For z $\ga$ 3.1, the continuum drop
indicating the absorption feature would move out of the IRS spectral
coverage.  It is not surprising, therefore, that the redshifts we have
identified fall in the range 1.6 $<$ z $<$ 3.1.  Within this range,
our initial results represent only a lower limit for the extent of
this obscured population.  This is primarily because of our
restriction to sources brighter than 0.75\,mJy at 24\,\ums, but also
because many sufficiently bright infrared sources with faint optical
counterparts (the 84 remaining sources having an optical
identification but with I $>24$ mag) were excluded from IRS
observation in the hope that redshifts for these very faint sources
might be determinable with ground-based spectroscopy.

Even though the sources in our sample are faint, their high redshifts
require luminosities that place them among the class of extreme or
``hyper-luminous" infrared galaxies. However, these sources have been
chosen because they are extremely faint in the optical, implying much
larger intrinsic extinction than in a typical ULIRG. For example, an
object having the infrared SED of Arp\,220 and sufficiently luminous
to have a 24\,\um flux density of 1\,mJy when at $z=2$ would have $I$
$<$23\,mag, making it too bright in the optical to be included in our
sample. (A source of the same luminosity as Arp\,220 if at $z=2$ would
be much too faint to be included in our infrared sample, having a 24
\um observed flux density of only 0.03 mJy.)  Thus, while our sources
have IRS spectra that can be well fit with known ULIRG templates in
the rest-frame mid-infrared, they are all much fainter optically, and
therefore apparently more heavily extincted, than low-redshift
ULIRGs. What is the nature of this obscured population?  Is the
mid-infrared luminosity of the hot dust produced primarily by an AGN
or by a starburst in our sample galaxies?  If IRAS F00183-7111 is
powered by an AGN, as suggested by \citet{spo04b}, then 13 out of 17
(76\%) of the sample are best fit with an AGN mid-IR template
(F00183-7111 or Mrk 231).  Four of the galaxies are best fit with a
starburst template (Arp\,220 or NGC\,7714).  Although the fraction of
light coming from an AGN or from a starburst in any ULIRG is
uncertain, these template fits suggest that most of the sources we
have found are AGN dominated -- i.e. steeply rising spectra with weak
PAH emission features and varying levels of silicate absorption.

Radio data can also aid in understanding the nature of these sources.
A well defined correlation exists between infrared and radio flux
densities for systems with no AGN contribution to the radio flux
density.  \citet{app04} show that this relation as described by the
parameter q = log$[$f$_{\nu}$(24 \ums)$/$f$_{\nu}$(20 cm)$]$ has q =
0.8 $\pm$ 0.3.  The Bo\"{o}tes survey area has been mapped as part of
a Westerbork 20\,cm survey \citep{dev02}.  Although the radio beam
($13\arcsec\times27\arcsec$) is much larger than even the MIPS 24\,\um
beam, we attempt source identifications by considering any 24 \um
source that has a quoted Westerbork source within $5\arcsec$ to be an
identification with this radio source.  With this criterion, five of
the infrared sources have radio counterparts.  For those that do not,
we take the upper limit to the 20\,cm flux density to be a 5\,$\sigma$
value of 0.20\,mJy.  The q values are listed in Table 1.  Except for
source \#13, all sources have values or limits for q very close to
those expected for starbursts or radio-quiet AGN.  Source \#13 does
have the excess radio emission which would characterize a radio-loud
AGN, and this source is best fit with the IRAS F00183-7111 template,
confirming at least for this case the validity of the AGN
interpretation.

To understand the contribution of this obscured population to the
luminosity density of the universe at high redshift, estimates of
bolometric luminosities are needed.  Our only way to estimate these is
to assume that mid-infrared luminosities scale to bolometric
luminosities the same as in the template sources which fit the
mid-infrared spectra.  Using this scaling, all sources in Table 1 for
which we have $z>1.7$ have implied infrared luminosities of
$6\times10^{12}$ -- $6\times10^{13}$L$_{\odot}$, with an average value
of $2.3\times10^{13}$L$_{\odot}$.  The largest existing samples of
dusty sources with comparable bolometric luminosities and redshifts
are the SCUBA or MAMBO submillimeter-selected galaxies
\citep{chap03}. Detection at 24\,\um of such sources brighter than a
few mJy at submillimeter wavelengths is common, but nearly all of
these galaxies are fainter at 24\,\um by a typical factor of 3--5 than
the sources reported here (e.g. \citet{ivi04}, \citet{ega04},
\citet{fra04}, and \citet{cha04}).  This implies a higher ratio of
mid-infrared to far-infrared luminosity in our sources than in the
sub-millimeter sources.  This indicates that our new sample seems to
represent sources that have hotter dust than typical
submillimeter-selected galaxies; this could be an additional component
of hot dust or an overall averaged dust temperature which is
greater. Either would be consistent with having AGN as the primary
luminosity source instead of the starbursts thought to power the
sub-millimeter galaxies. \citet{chap04} utilized optically determined
redshifts to identify a population of optically faint radio galaxies
($R$ $>$ 23.5\,mag) at z $\sim$ 2 which are not detected as
submillimeter sources, and half of this sample has optical indicators
of an AGN. Although none of the sources were observed in the infrared,
\citet{chap04} suggest from the absence of submillimeter detections
that this sample is also characterized by hot dust, so it may
represent a similar population to our sample.

\section{Summary}

The NOAO Deep Wide-Field Survey area in Bo\"{o}tes has been surveyed
at $24\,\mu$m with the MIPS instrument on SST, and 31 optically faint
or invisible sources have been observed with the IRS instrument. Our
first results indicate that sources chosen solely on the basis of
extreme 24 \um to $I$ band flux density ratios, $\nu$f$_{\nu}$(24
\ums)$/ \nu$f$_{\nu}$($I$) $\ga$ 100, are heavily selected in favor of
redshifts exceeding 1.5 and that the majority of such sources have
redshifts that are measurable with IRS based on a strong silicate
absorption feature in the mid-infrared.  Sixteen of the 31 objects
selected for the initial IRS observations are found to have
spectroscopic redshifts 1.7 $<$ z $<$ 2.8, based on fits of redshifted
IRS spectral templates for local ULIRGS, and 9 of these have no
optical counterparts.  The template spectra and radio survey data
suggest that this population is dominated by AGN-powered, radio-quiet
ULIRGs.  If bolometric luminosities scale with mid-infrared
luminosities as in the templates, these newly discovered sources have
bolometric luminosities between $6\times 10^{12}$L$_{\odot}$ and
$6\times 10^{13}$L$_{\odot}$.

\acknowledgments

We thank many others on the IRS, MIPS, and Bo\"otes Survey teams for
contributing to this effort, especially including D.  Barry,
D. Devost, P. Hall, G. Sloan, H. Dole, P. Appleton, and C.  Bian.
This work is based in part on observations made with the Spitzer Space
Telescope, which is operated by the Jet Propulsion Laboratory,
California Institute of Technology under NASA contract 1407. Support
for this work was provided by NASA through Contract Number 1257184
issued by JPL/Caltech.  This work used data provided by the
NDWFS, which is supported by NOAO. NOAO is operated by AURA, Inc.,
under an agreement with the NSF.

\clearpage




\begin{deluxetable}{lcccccrccrrc} 
\tabletypesize{\footnotesize}
\tablecaption{Redshifts and Source Characteristics}
\tablewidth{0pc}
\tablehead{
  \colhead{Source} & \colhead{Name\tablenotemark{a}} &
  \colhead{f$_{\nu}$ 24\ums} & \colhead{$B_W$}
  & \colhead{$R$}  & \colhead{$I$}  & \colhead{time\tablenotemark{b}}& \colhead{z} & \colhead{Template} & \colhead{IR/opt\tablenotemark{c}}& \colhead{q} & \colhead{L$_{ir}$\tablenotemark{d}}\\
    &  &\colhead{(mJy)} & \colhead{(mag)} & \colhead{(mag)} & \colhead{(mag)} & \colhead{(sec)}& & & & & ($\times10^{13}$\,L$_{\odot}$)}
\startdata
1.  & SST24 J142958.33+322615.4 &   1.24 & 25.6    & 25.1    & 24.4    & 1200 & 2.64$\pm$.25 & F00183-7111 &    110 & $>$0.8 & 1.8\\
2.  & SST24 J142653.23+330220.7 &   0.89 & \nodata & \nodata & 24.7    & 1680 & 1.86$\pm$.07 & NGC 7714    &    100 & $>$0.8 & 0.9\\
3.  & SST24 J143447.70+330230.6 &   1.81 & \nodata & \nodata & \nodata &  960 & 1.78$\pm$.30 & Mrk 231     & $>$270 &    0.8 & 0.9\\
4.  & SST24 J143523.99+330706.8 &   1.08 & 26.7    & 24.6    & 23.5    & 1440 & 2.59$\pm$.34 & F00183-7111 &     40 &    0.8 & 1.5\\
5.  & SST24 J142804.12+332135.2 &   0.87 & \nodata & \nodata & \nodata & 1680 & 2.34$\pm$.28 & F00183-7111 & $>$130 & $>$0.6 & 0.9\\
6.  & SST24 J143358.00+332607.1 &   1.03 & \nodata & \nodata & \nodata & 1440 & 1.96$\pm$.34 & Mrk 231     & $>$150 & $>$0.7 & 0.6\\
7.  & SST24 J143251.82+333536.3 &   0.78 & 26.2    & \nodata & 24.3    & 1920 & 1.78$\pm$.14 & Arp 220     &     60 & $>$0.8 & 2.8\\
8.  & SST24 J143539.34+334159.1 &   2.65 & 26.2    & 25.4    & 24.5    &  720 & 2.62$\pm$.26 & F00183-7111 &    250 & $>$1.1 & 3.8\\
9.  & SST24 J143001.91+334538.4 &   3.83 & \nodata & \nodata & \nodata &  480 & 2.46$\pm$.20 & F00183-7111 & $>$570 &    1.0 & 4.6\\
10. & SST24 J143520.75+340418.2 &   1.45 &  25.5   & 25.3    & 24.3    & 1200 & 2.08$\pm$.21 & Mrk 231     &    110 & $>$0.8 & 1.0\\
11. & SST24 J143242.51+342232.1 &   0.87 & \nodata & \nodata & \nodata & 1680 & 0.70$\pm$.24 & Mrk 231     & $>$130 & $>$0.6 & 3.0\\
12. & SST24 J142626.49+344731.2 &   1.12 & \nodata & \nodata & \nodata & 1440 & 2.13$\pm$.09 & Arp 220     & $>$170 &    0.8 & 5.7\\
13. & SST24 J143644.22+350627.4 &   2.30 &  24.7   & 23.7    & 23.4    &  720 & 1.95$\pm$.17 & F00183-7111 &     80 &   -0.3 & 1.8\\
14. & SST24 J142538.22+351855.2 &   0.79 & \nodata & \nodata & \nodata & 1920 & 2.26$\pm$.11 & Arp 220     & $>$120 & $>$0.6 & 5.3\\
15. & SST24 J142645.71+351901.4 &   1.05 & \nodata & \nodata & \nodata & 1440 & 1.75$\pm$.21 & Mrk 231     & $>$160 & $>$0.8 & 0.5\\
16. & SST24 J142924.83+353320.3 &   1.04 & \nodata & \nodata & \nodata & 1440 & 2.73$\pm$.19 & F00183-7111 & $>$150 & $>$0.7 & 1.6\\
17. & SST24 J143504.12+354743.2 &   1.23 & \nodata & \nodata & \nodata & 1200 & 2.13$\pm$.17 & Mrk 231     & $>$180 & $>$0.8 & 0.9\\
\enddata

\tablenotetext{a}{SST24 source name derives from discovery with the
  MIPS 24$\mu$m images; coordinates listed are J2000 24$\mu$m
  positions with typical 3\,$\sigma$ uncertainty of $\pm$ 1.2\arcsec;
  sources with an optical counterpart will also appear in NDWFS
  catalogs with prefix NDWFS and the optical source position; optical
  magnitudes are Mag-Auto from NDWFS Data Release Three, available at
  http://www.noao.edu/noao/noaodeep/).}  \tablenotetext{b}{Integration
  time for each order of Long Low spectrum; integration time in Short
  Low order 1 is 480 s in all cases.} \tablenotetext{c}{IR/opt $=
  \nu$f$_{\nu}$(24 \ums)$/ \nu$f$_{\nu}$($I$).  Limits are based on
  assumed $I$ $>$ 25.}  \tablenotetext{d}{The instrinsic infrared
  luminosity of each source, L$_{ir}$=L(8-1000$\mu$m), is estimated by
  scaling the known L$_{ir}$ of the best template fit to the measured
  MIPS 24$\mu$m flux density after redshifting the template SED to its
  measured z.}
\end{deluxetable}

%
%

\clearpage

\begin{figure}
\figurenum{1}
\includegraphics[scale=0.5,angle=0]{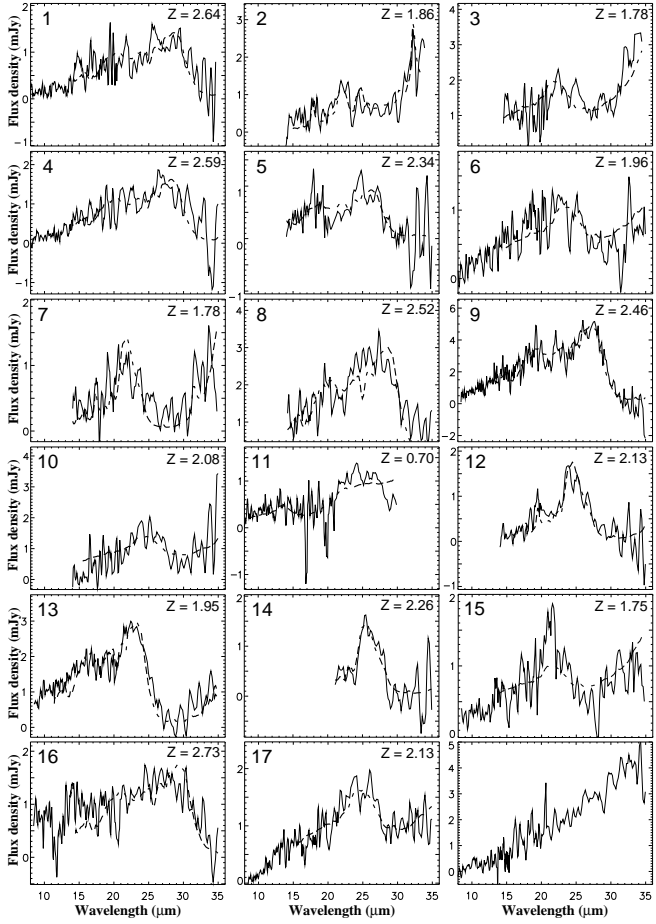}
\caption{ Observed spectra smoothed to approximate resolution of individual IRS orders (solid line), best template fit (dashed line),
  observed flux densities and redshift for all 17 sources in Table 1.
  Spectra truncated below 15 \um had no detectable signal in SL1;
  spectrum \#14 also had no signal in LL2.  Final panel is a source not
  listed in Table 1 to illustrate a spectrum without sufficient
  features for a redshift measurement. }
\end{figure}

\clearpage
\begin{figure}
 \figurenum{2}
\includegraphics[scale=0.5]{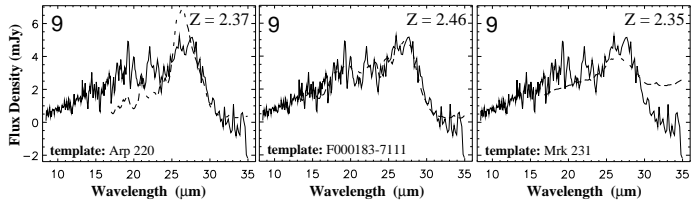}
\caption{Spectra of source \#9 showing alternative template fits with
  Arp\,220, absorbed ULIRG IRAS F000183-7111, and Mrk\,231. Range of z values
  illustrates uncertainty in z that arises from different template
  assumptions.}
\end{figure}

\end{document}